\documentclass[conference,a4paper]{IEEEtran}
\usepackage{graphicx}
\usepackage{subcaption}
\usepackage{amsmath}
\usepackage{color}
\usepackage{url}
\usepackage{float}
\usepackage[normalem]{ulem}

\title{Short-Form Video Viewing Behavior Analysis and Multi-Step Viewing Time Prediction}
\author{

\IEEEauthorblockN{Vu Thi Hai Yen$^1$, Duc V. Nguyen$^2$, Cao Anh Minh Huy$^1$, Truong Thu Huong$^1$}
\IEEEauthorblockA{
$^1$\textit{School of Electrical and Electronic Engineering, Hanoi University of Science and Technology, Vietnam} \\
$^2$\textit{Faculty of Engineering, Tohoku Institute of Technology, Japan}}
}
\date{}

\begin{document}
\maketitle
\begin{abstract}
    Short-form videos have become one of the most popular user-generated content formats nowadays. Popular short-video platforms use a simple streaming approach that preloads one or more videos in the recommendation list in advance. However, this approach results in significant data wastage, as a large portion of the downloaded video data is not used due to the user's early skip behavior. To address this problem, the chunk-based preloading approach has been proposed, where videos are divided into chunks, and preloading is performed in a chunk-based manner to reduce data wastage. To optimize chunk-based preloading, it is important to understand the user's viewing behavior in short-form video streaming. In this paper, we conduct a measurement study to construct a user behavior dataset that contains users' viewing times of one hundred short videos of various categories. Using the dataset, we evaluate the performance of standard time-series forecasting algorithms for predicting user viewing time in short-form video streaming. Our evaluation results show that Auto-ARIMA generally achieves the lowest and most stable forecasting errors across most experimental settings. The remaining methods, including AR, LR, SVR, and DTR, tend to produce higher errors and exhibit lower stability in many cases. The dataset is made publicly available at https://nvduc.github.io/shortvideodataset. 
\end{abstract}
\section{Introduction}
% characteristics of short-video
Short videos have become the most popular online video format, being used extensively for entertainment, marketing, and education. Short video platforms such as TikTok, YouTube Shorts, and Instagram Reel are attracting billions of users, with millions of new short videos uploaded monthly~\cite{shortvideostat2025}. Short videos are typically less than a few minutes long. Most short videos are in vertical format and are primarily viewed on smartphones. Especially, users frequently swipe to watch many short videos during a streaming session~\cite{Zhang2022}.

% data wastage problem
In a short video streaming system, the system recommends videos that are relevant to the user based on factors such as the user's preferences and past viewing data~\cite{Tiktok_recommend_alg}. The user agent (e.g., Mobile App) downloads the recommended videos from dedicated cloud servers and displays the videos on the user device. The user can switch to the next/previous video in the recommended list at any time by swiping up/down. To ensure a smooth viewing experience, short video platforms employ a simple streaming approach that preloads several videos in the recommended list in advance~\cite{SwipeAlong}. The preloaded videos are stored in the client-side buffer. Video playback is switched to the next/previous video in the recommended list upon user swiping event. This approach, however, causes significant data wastage because most of the short videos are skipped very early during playback~\cite{Zhang2022}. For example, if a user only watches the first 10 seconds of a 50-second-long video, then 80\% of the downloaded data is wasted.

% chunk-based preloading solution
To reduce data wastage without degrading the user's Quality of Experience, the chunk-based preloading approach has been proposed~\cite{nguyen2022,Phong2023,DeLoad}. In the chunk-based preloading approach, short videos are divided into small chunks with a short playback duration (e.g., 1 s) in advance. Instead of preloading the entire video, the user agent maintains a preloading buffer with a pre-calculated number of chunks (e.g., 5 chunks). The number of preloaded chunks is mainly decided based on network conditions (e.g., throughput) and user viewing behavior (i.e., viewing time). By choosing a suitable preloading size, the chunk-based preloading approach can significantly reduce the amount of data wastage without affecting the user's viewing experience~\cite{nguyen2022}. Each video chunk can be further encoded into multiple bitrate versions to cope with varying network conditions~\cite{Phong2023}.

% the need to have a user viewing time dataset (research gap)
 Both network-level and user-level information is required to design and evaluate chunk-based preloading solutions. While network-level data, such as bandwidth, have been collected and made publicly available~\cite{dataset_3G,datast_4G, dataset_5G}, user-level data in short video streaming is very limited. Although there have been several prior works on collecting user behavior data on short-form video platforms, previously collected datasets are either not publicly available~\cite{Zhang2022,dashlet} or not suitable for streaming tasks~\cite{dataset_KuaiRec, dataset_reasoner, dataset_TsinghuaUni,dataset_vRetention}. To fill in this gap, we develop a Web-based tool that allows researchers/practitioners to collect user swipe data in short video streaming. Using the developed tool, we conduct a measurement study to collect the users' viewing times in short video streaming sessions. The collected data consist of viewing times for one-hundred short videos, each with 50 users. We further evaluate the performance of the viewing time prediction algorithms.

The remainder of this paper is organized as follows. Related works are reviewed in Section~\ref{sec:related_work}. The construction of the user behavior dataset is described in Section~\ref{sec:dataset}. The evaluation of the viewing time prediction algorithm is shown in Section~\ref{sec:view_time_prediction}. Finally, the paper is concluded in Section~\ref{sec:conclusion}.

\section{Related Work}\label{sec:related_work}
% Measurement study, highlighting the lack of a high-quality dataset for short video streaming 
So far, content characteristics and user viewing behavior on short video services have been studied, showing that users have a very short attention span. In~\cite{Zhang2022}, it is found that more than 40\% of short videos on a commercial short video service have a playback percentage less than 25\%. Study on a popular news application found that only 10\% of videos are viewed in their entirety with an average playback percentage of 57\%~\cite{Measurement_Study_2025}. An analysis of retention curves of over 200,000 videos shows that YouTube short videos exhibit early abandonment with average playback percentage lower than 50\%~\cite{dataset_vRetention}. Meanwhile, popular short video platforms including YouTube Shorts, TikTok, Instagram Reel, and Facebook Watch employ a simple streaming approach that preloads videos in the recommendation list~\cite{SwipeAlong}. The number of preloaded videos varies across short video services, ranging from one video to up to 22 videos~\cite{SwipeAlong}. 

% Overview of chunk-based preloading methods, highlighting the need for user viewing time dataset
To reduce data wastage without affecting users' QoE, both heuristic-based (e.g.,~\cite{nguyen2022,Phong2023,dashlet}) and learning-based chunk-based preloading solutions (e.g.,~\cite{LiveClip,Gamora,DeLoad}) have been developed. In heuristic solutions, the optimal preloading size and bitrate of individual videos are determined based on the user's past viewing behavior and network conditions~\cite{nguyen2022,Phong2023}. In a learning-based approach, users' past viewing behaviors, network conditions, and video metadata are used to learn an optimal preloading policy using reinforcement learning~(e.g., ~\cite{Gamora}).

For performance evaluation, prior works mainly use the dataset provided in~\cite{MM22challenge}. However, this dataset contains only seven videos, and user viewing time per video in a streaming session is sampled from pre-calculated swipe distributions~\cite{MM22challenge}. Real user swipe traces are collected and used for performance evaluation of preloading solutions in~\cite{dashlet}. However, the collected data in~\cite{dashlet} is not made publicly available. Some large-scale public datasets containing short video user behavior have been collected~\cite{dataset_TsinghuaUni, dataset_KuaiRec,dataset_tenrec}. Yet, these datasets are mainly for the recommendation task.

\begin{figure}[t]
    \centering
\includegraphics[width=\columnwidth]{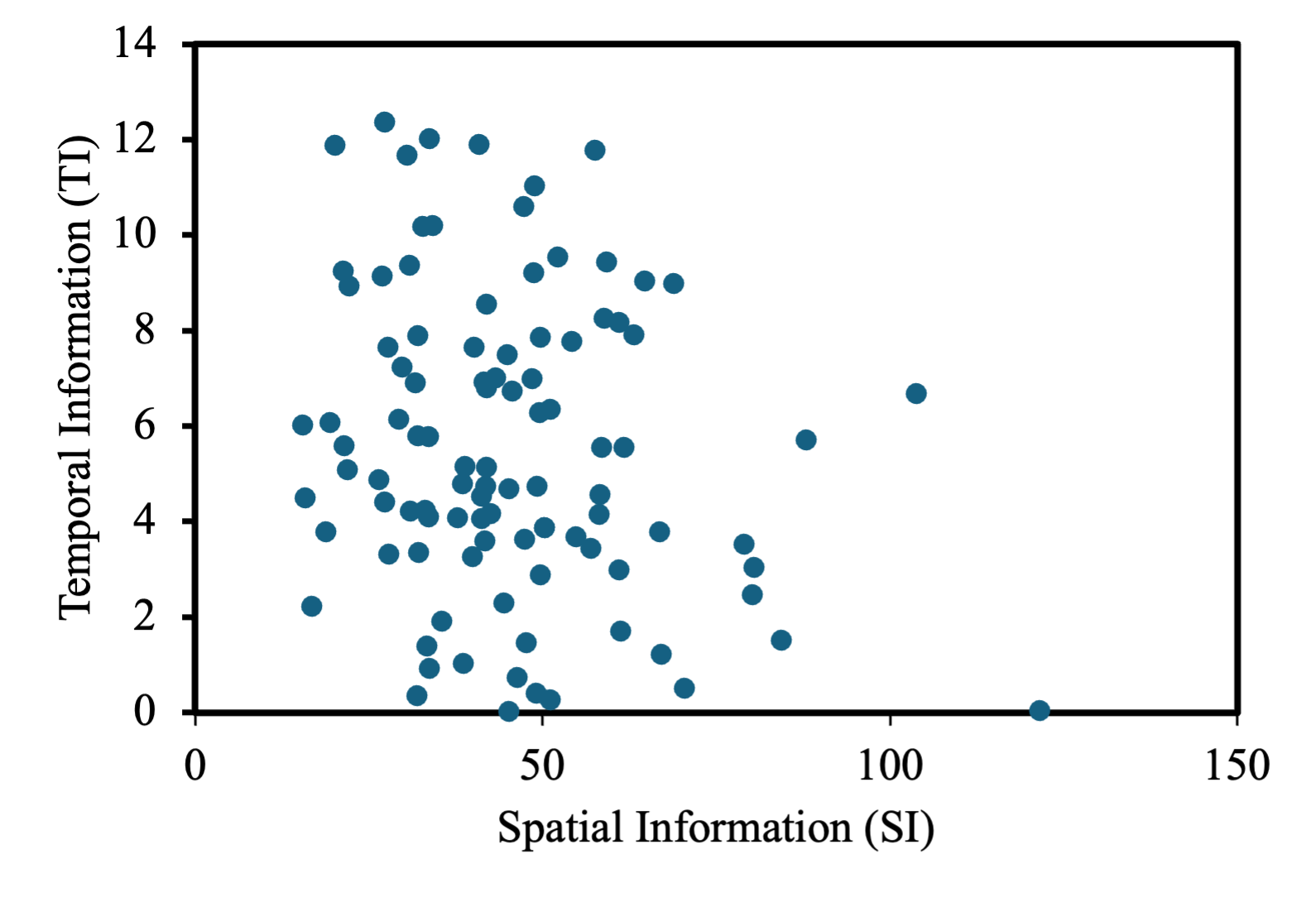}
    \caption{Spatial and temporal information values of short videos used in our experiment.}
    \label{fig:siti}
\end{figure}
\begin{figure}[t]
    \centering
    \begin{subfigure}{0.45\linewidth}
        \centering
        \includegraphics[width=\linewidth]{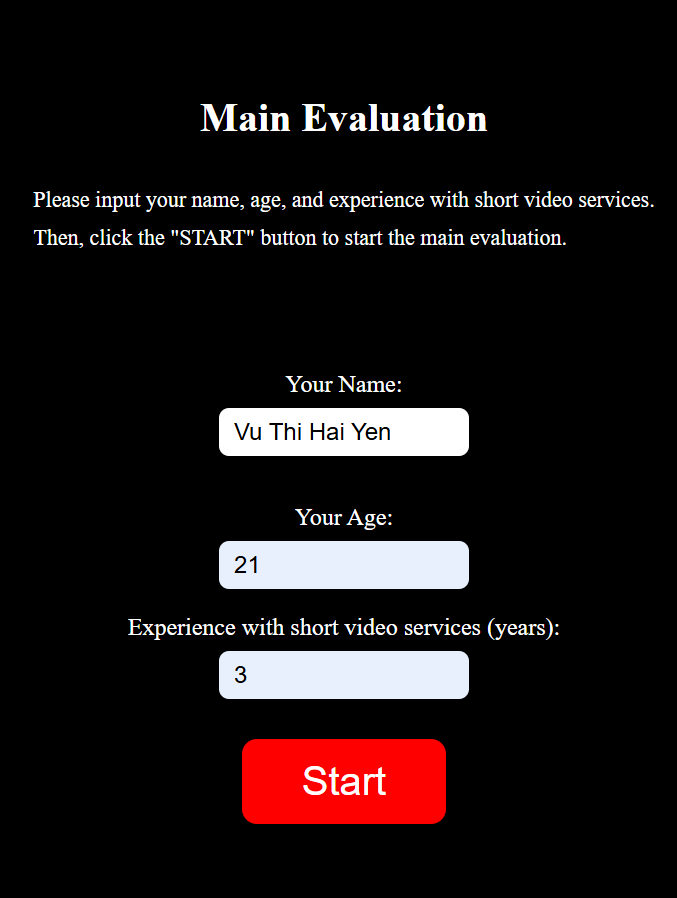}
        \caption{Application interface}
        \label{fig:app_ui}
    \end{subfigure}
    \hfill
    \begin{subfigure}{0.45\linewidth}
        \centering
        \includegraphics[width=\linewidth]{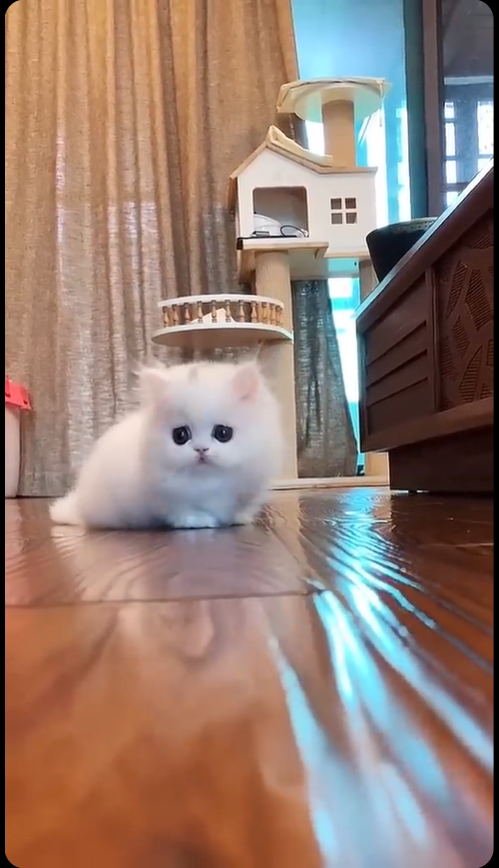}
        \caption{Main screen}
        \label{fig:view_log}
    \end{subfigure}
    \caption{Screenshots of the custom-developed short video streaming application.}
    \label{fig:playback_app}
\end{figure}
\section{User Viewing Time Data  Collection}\label{sec:dataset}
\subsection{Data Collection}
Viewing time data is necessary to evaluate the performance of viewing time prediction algorithms. However, session-based viewing time data is very limited in the literature. To address this problem, we conduct a measurement study to collect user viewing times in short video streaming. First, we select 100 short videos of various categories from YouTube Shorts. The selected videos span 20 distinct content categories to ensure diversity in user viewing behavior. These categories include educational content (Maths, Physics, Chemistry, Biology, and Coding), entertainment and media (Singer, Song, Drama, and VTV), gaming and e-sports (TFT and Arena of Valor), as well as lifestyle and social content (Food Review, Psychology, and Flexing). In addition, the dataset covers digital and creative content (VTuber) and informational or exploratory videos (Nature, Animal, Discovery, War, and HUST-related content). Each category contains five videos with varying durations, which allows the dataset to capture differences in user attention and viewing time even within the same content type. This balanced categorization helps mitigate category bias and supports a more accurate evaluation of user behavior in realistic short video streaming scenarios.
. The videos' durations range from 4 to 81 seconds, with an average duration of 34.95 seconds. All videos have a resolution of 720p. The spatial information (SI) and temporal information (TI) of the considered videos are shown in Fig.~\ref{fig:siti}.

Videos are presented to users using a Web-based short video streaming player that resembles the interface of YouTube Shorts, as shown in Fig.~\ref{fig:playback_app}. The videos are divided into five sessions, each consisting of twenty videos. Users are free to swipe to watch the next video at any time during a session. There is a 5-minute rest period between sessions. The experiment is conducted using a crowdsourcing approach where each participant uses their own device to participate in the experiment. In total,  fifty people aged between 18 and 30 participated in our experiment.

\subsection{Data Analysis}

The watch-time retention ratio is defined as the fraction of the video duration that a user watches during a viewing
session. Specifically, for a viewing session associated with
video $t$ and user $u$, the retention ratio $r_{u,t}$ is computed as
\begin{equation}
r_{u,t} = \frac{y_{u,t}}{L_t},
\end{equation}
where $y_{u,t}$ denotes the actual watch time of user $u$ on
video $t$, and $L_t$ represents the total duration of the video.
The retention ratio is bounded in the range $[0,1]$, with
$r_{u,t}=1$, indicating that the video is watched in its entirety.

\begin{figure}[t]
    \centering
    \includegraphics[width=\columnwidth]{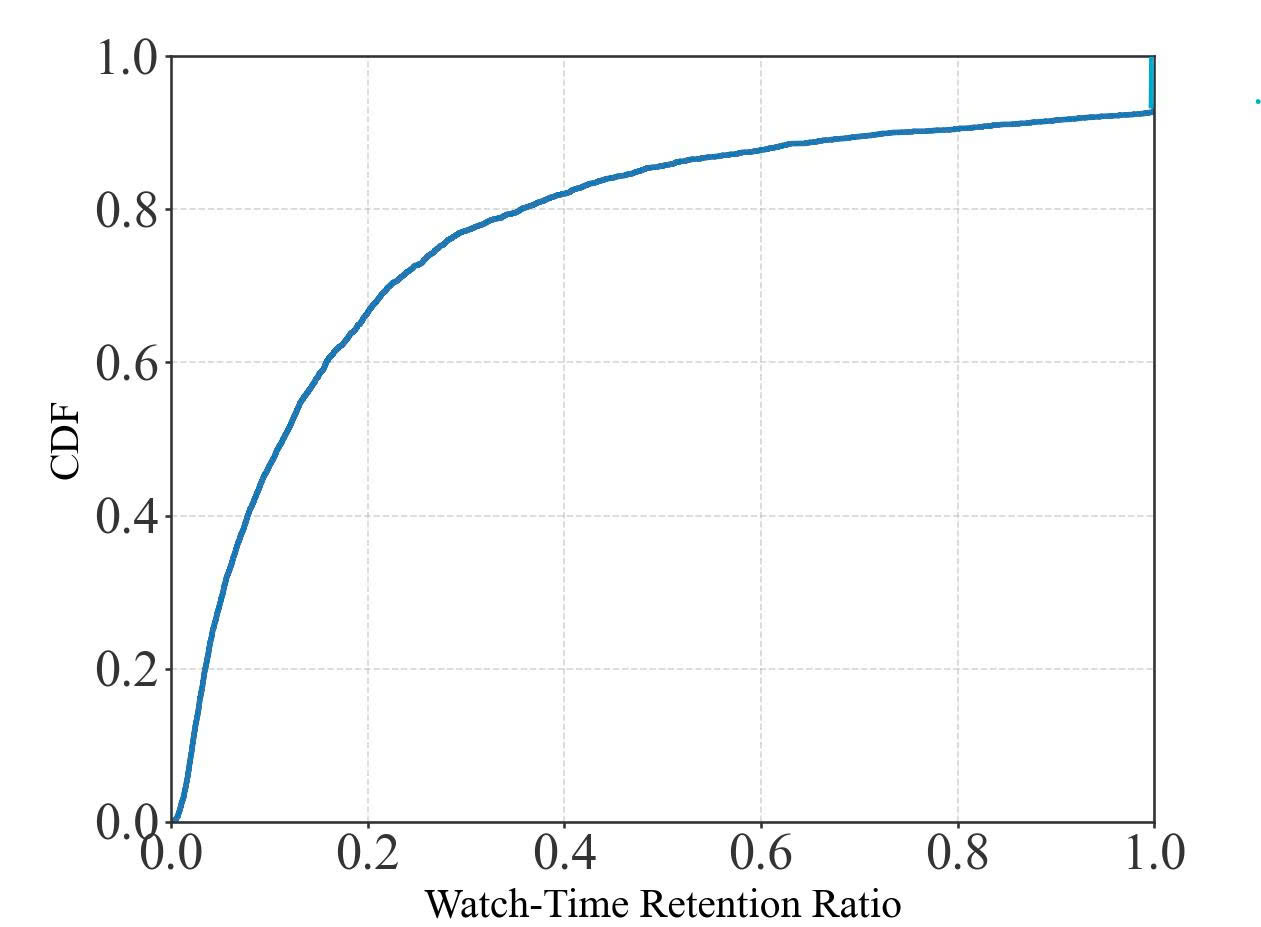}
    \caption{CDF of watch-time retention ratio aggregated over all viewing sessions
from 50 users}
    \label{fig:cdf_user}
\end{figure}
\begin{figure}[t]
    \centering
    \includegraphics[width=\columnwidth]{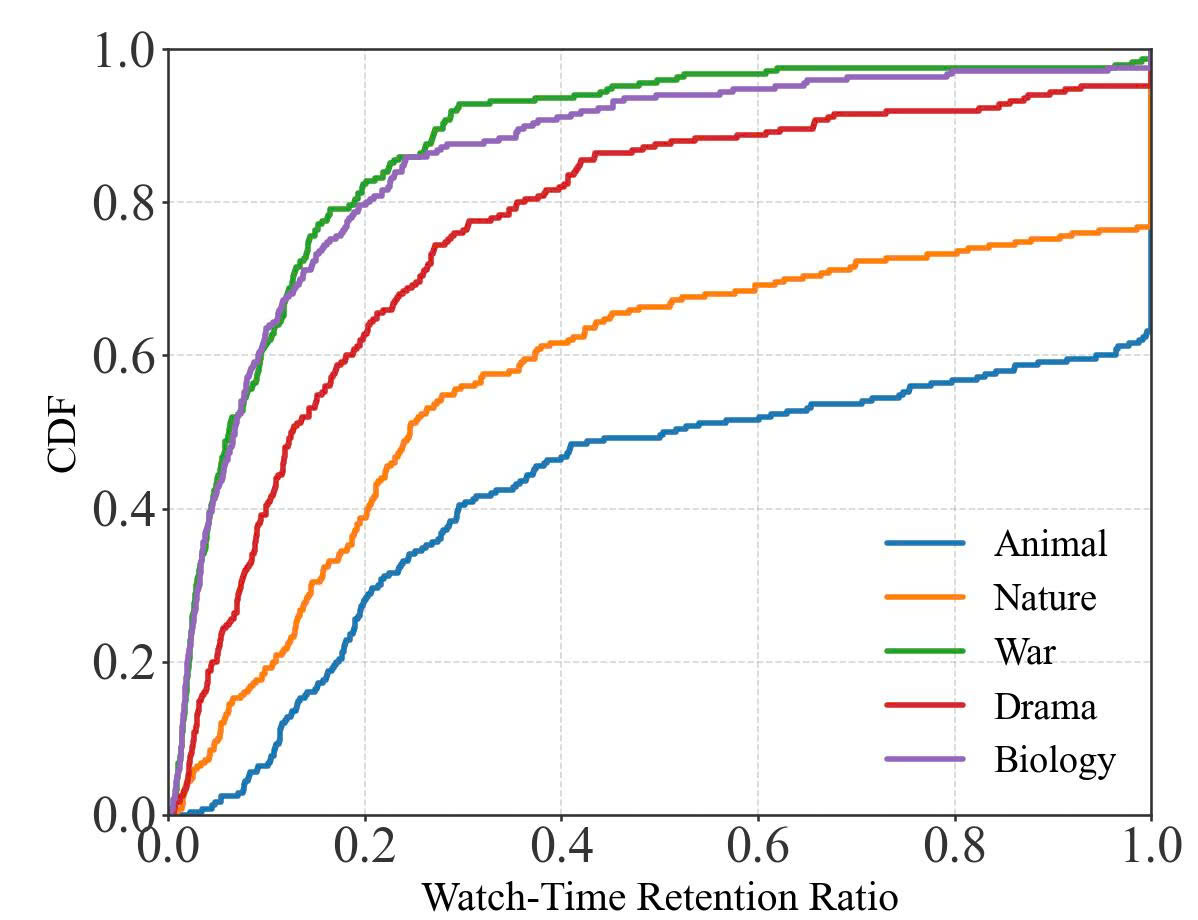}
    \caption{CDF of  watch-time retention ratio across different video categories}
    \label{fig:cdf_category}
\end{figure}
Fig.~\ref{fig:cdf_user} presents the cumulative distribution function (CDF)
of watch-time retention ratio aggregated over all viewing sessions from
50 users. The CDF exhibits a steep increase in the low retention region,
indicating that a large proportion of viewing sessions terminate at an
early stage of video playback. Specifically, approximately 70\% of viewing sessions end before reaching 20\% of the video duration, highlighting a pronounced early abandonment phenomenon that is typical of short-form video consumption. As the retention ratio increases, the CDF gradually flattens and
approaches saturation, suggesting that long viewing sessions—where users watch most or all of the video—occur with significantly lower probability. Overall, the aggregated CDF reveals a highly left-skewed distribution of viewing behavior, dominated by short-duration engagement.

Fig.~\ref{fig:cdf_category} presents the cumulative distribution functions (CDFs) of the watch time retention ratio across different video content categories. Although all CDF curves exhibit a similar overall shape characterized by a steep rise in the early stage, notable differences can be observed in their characteristic values and tail behaviors, indicating heterogeneous user
engagement across content categories. Among the examined categories, the \emph{Animal} category achieves the highest median retention ratio, suggesting stronger initial attractiveness and higher viewer engagement. In contrast, categories such as \emph{War} and \emph{Biology} exhibit noticeably lower median values, reflecting a higher tendency for early abandonment. These discrepancies indicate that viewers’ decisions to continue watching are strongly influenced by the immediate accessibility and appeal of the content. The differences among categories are further amplified in the tail regions of the distributions. More than 20\% of viewers continue watching videos in the \emph{Animal} and \emph{Nature} categories beyond a retention ratio of 0.2, whereas this proportion drops to approximately 14\% for the \emph{Drama} and \emph{War} categories. This observation suggests that entertainment-oriented and visually rich content is more effective at sustaining user attention over longer viewing durations.

\textcolor{black}{ To evaluate the stationarity characteristics of video watch-time sequences, this study employs the Augmented Dickey--Fuller (ADF) test, a widely used statistical method for detecting the presence of a unit root in time-series data~\cite{ADFtest}. A unit root can be understood as a condition in which the current value of a series depends strongly on its past values, causing the series to drift over time rather than fluctuate around a stable level. When a unit root is present, shocks or changes in the past may have persistent effects on future values, leading to non-stationary behavior. In the ADF framework, the null hypothesis ($H_0$) states that the time series contains a unit root and is therefore non-stationary, while the alternative hypothesis ($H_1$) states that the series is stationary. Rejection of the null hypothesis is determined based on the p-value, which represents the probability of observing the test statistic under the assumption that $H_0$ is true. Smaller p-values provide stronger evidence against the presence of a unit root. The ADF test is applied independently to the watch-time sequences of 50 users, each consisting of 100 consecutive observations. The results show an average p-value of 0.029 and a median p-value close to zero, indicating that the majority of the sequences reject the unit-root hypothesis at the 5\% significance level. Specifically, approximately 92\% of the sequences are identified as stationary, while 8\% remain non-stationary. Therefore, the watch-time data can be considered predominantly stationary across users, with a small proportion exhibiting non-stationary behavior.
} 
\section{Viewing Time Prediction}\label{sec:view_time_prediction}

\subsection{Problem Setup}
Consider a scenario in which a user watches a sequence of short videos on a personal device (e.g., a smartphone). For brevity, we omit the user index $u$ when describing the prediction models. Let $y_t$ denote the viewing time of the $t$-th video in the sequence, where $t = 1, 2, \ldots$.
The viewing times of the last $N$ videos, denoted by $\{y_t\}_{t=1}^{N}$, are used as training data.
The objective is to predict the viewing times of the next $H$ videos, denoted by $\{\hat{y}_t\}_{t=N+1}^{N+H}$,
where $H$ represents the prediction horizon.
This formulation corresponds to a session-based time-series forecasting problem, in which future viewing behavior is inferred from historical viewing patterns.

\subsection{Prediction Models}
In this study, five representative time-series prediction methods are employed to forecast future short-video viewing times,
including Linear Regression (LR), AutoRegression (AR), Auto-ARIMA, Support Vector Regression (SVR),
and Decision Tree Regression (DTR).
All models are trained using the same historical data $\{y_t\}_{t=1}^{N}$ and evaluated under identical
forecasting settings to ensure a fair comparison.

\subsubsection{Linear Regression (LR)}
Linear Regression is used as a baseline model to capture the global temporal trend of the viewing time sequence.
The model is defined as
\begin{equation}
\hat{y}_t = \beta_0 + \beta_1 t,
\end{equation}
where $\beta_0$ and $\beta_1$ are regression coefficients estimated from the training data.
Future viewing times are predicted by extrapolating the learned linear trend to time indices $N < t \le N+H$.
Unlike other models, LR does not explicitly utilize past viewing times as input features,
but instead models the overall temporal evolution of the sequence.

\subsubsection{AutoRegression (AR)}
The AutoRegression (AR) model predicts future values as a linear combination of past observations.
An AR model of order $p$ is expressed as
\begin{equation}
\hat{y}_t = \sum_{k=1}^{p} \phi_k y_{t-k},
\end{equation}
where $\phi_k$ are autoregressive coefficients estimated from the training data.
The AR model captures short-term temporal dependencies and is typically applied to \textcolor{black}{stationary} time-series data. In this study, the lag order is set to $p = \min(4, N)$. Since the time series within each session is typically short, only a small number of recent observations is needed for prediction. Limiting the lag order to at most four prevents the model from becoming overly complex and reduces the risk of overfitting, while ensuring that $p \leq N$ when the training sample size is small.

\subsubsection{Auto-ARIMA}

Auto-ARIMA extends the AR model by incorporating differencing and moving-average components to handle non-stationary time series. As introduced earlier, $y_t$ denotes the viewing time of the $t$-th video in a session, where $t = 1,2,\ldots$. The training set consists of $\{y_t\}_{t=1}^{N}$, and the objective is to predict $\hat{y}_t$ for $N < t \leq N+H$.
First, the series is differenced $d$ times, denoted by $\Delta^d y_t$. The $d$-th order differencing operator can be expressed in general form as
\begin{equation}
\Delta^d y_t
=
\sum_{k=0}^{d}
(-1)^k
\binom{d}{k}
y_{t-k}.
\end{equation}
After differencing, the ARIMA$(p,d,q)$ model is fitted to the differenced series $\Delta^d y_t$. The predicted value in the original scale is obtained through inverse differencing and is given by
\begin{equation}
\hat{y}_t
=
\Delta^{-d}
\left(
\sum_{k=1}^{p} \phi_k \Delta^d y_{t-k}
+
\sum_{j=1}^{q} \theta_j \varepsilon_{t-j}
\right),
\label{eq:arima}
\end{equation}
where $\phi_k$ and $\theta_j$ denote the autoregressive and moving-average coefficients, respectively, and $\varepsilon_t$ represents white noise. The operator $\Delta^{-d}$ is the inverse of the $d$-th order differencing operator and is used to restore the predicted values to the original scale. For the viewing-time sequence $y_t$. The parameters $(p,d,q)$ are automatically selected using data-driven model selection criteria.

\subsubsection{Support Vector Regression (SVR)}
Support Vector Regression (SVR) aims to learn a nonlinear regression function
whose prediction errors are bounded within a predefined margin $\varepsilon$.
For each time index $t$, the input feature vector is constructed from the past $p$ observations as
\begin{equation}
\mathbf{x}_t = [y_{t-1}, y_{t-2}, \ldots, y_{t-p}]^{\top},
\end{equation}  This vector is used to generate the prediction $\hat{y}_t$ for both the training phase ($p < t \le N$) and the forecasting phase ($N < t \le N+H$). The SVR prediction function in kernel form is given by
\begin{equation}
\hat{y}_t = \sum_{i=p+1}^{N} (\alpha_i - \alpha_i^{*}) K(\mathbf{x}_i, \mathbf{x}_t) + b,
\end{equation}
where $\alpha_i$ and $\alpha_i^{*}$ are learned dual variables, $b$ is the bias term,
and $K(\cdot,\cdot)$ denotes the kernel function.
In this work, the radial basis function (RBF) kernel is adopted:
\begin{equation}
K(\mathbf{x}_i, \mathbf{x}_t) =
\exp\left(-\gamma \lVert \mathbf{x}_i - \mathbf{x}_t \rVert^2\right),
\end{equation}
where $\gamma$ controls the kernel width.
Through nonlinear kernel mappings, SVR is capable of modeling complex temporal dependencies
in viewing time sequences.

\begin{figure*}[!t]
    \centering
    \includegraphics[width=0.24\textwidth]{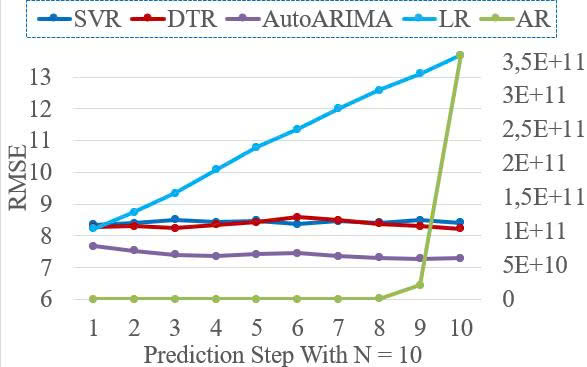}
    \includegraphics[width=0.24\textwidth]{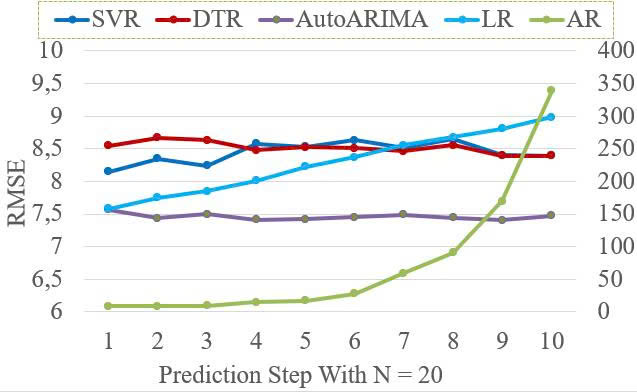}
    \includegraphics[width=0.24\textwidth]{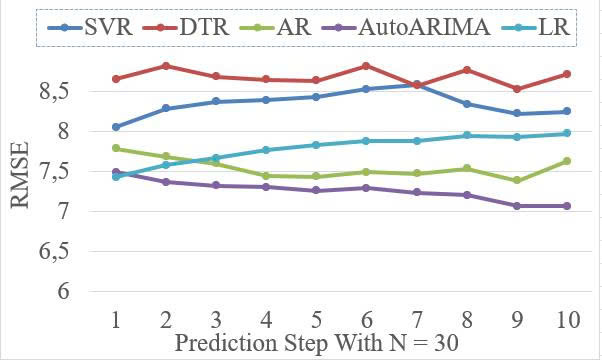}
    \includegraphics[width=0.24\textwidth]{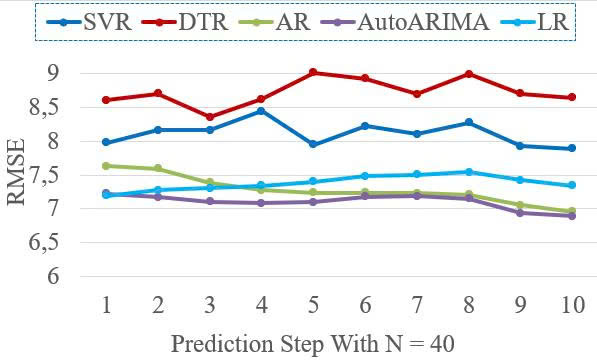}

    % \vspace{2mm}
    \hfill

    \includegraphics[width=0.24\textwidth]{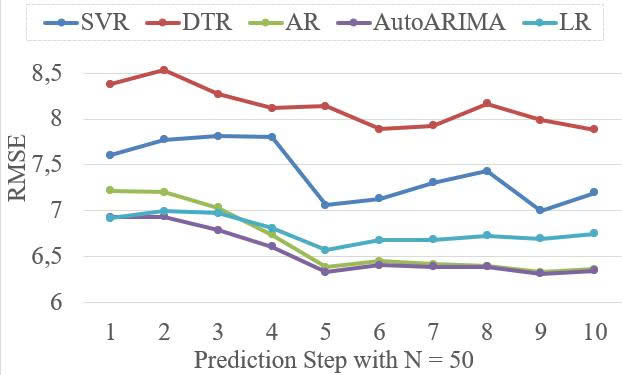}
    \includegraphics[width=0.24\textwidth]{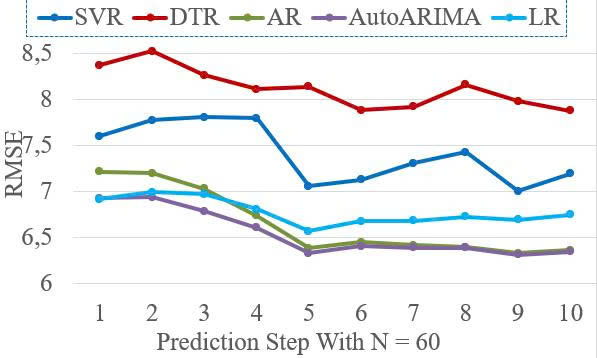}
    \includegraphics[width=0.24\textwidth]{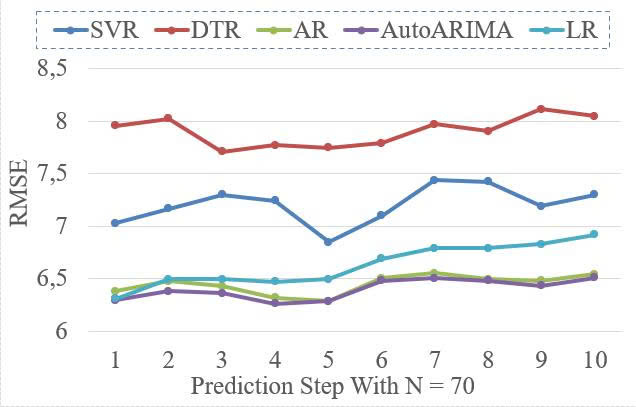}
    \includegraphics[width=0.24\textwidth]{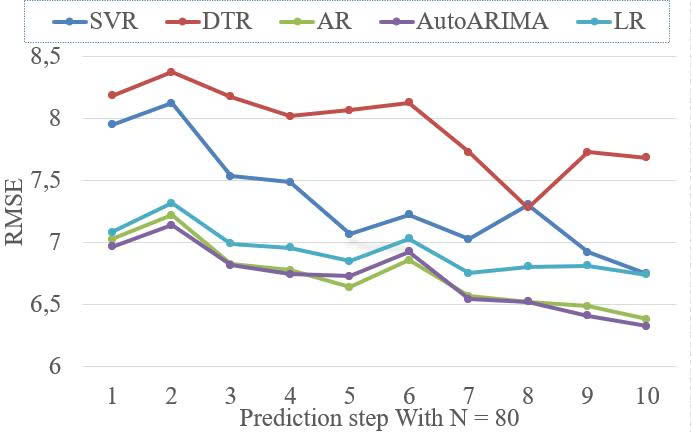}
    \caption{Comparison of the RMSE of different methods for $N=10$ to $N=80$.} 
    \label{fig:rmse_all}
\end{figure*}
\subsubsection{Decision Tree Regression (DTR)}
Decision Tree Regression (DTR) is a nonlinear forecasting model that predicts future values by recursively partitioning the input feature space defined by historical observations. Given the input vector $x_t$ (defined similarly to the SVR model for $p < t \le N+H$), the prediction can be written as:
\begin{equation}
\hat{y}_t = \sum_{m=1}^{M} c_m I(x_t \in R_m), \quad 
\end{equation}
where $R_m$ denotes the $m$-th leaf region of the decision tree, $c_m$ is the constant prediction value assigned to that region, and $I(\cdot)$ is the indicator function. DTR can capture nonlinear and non-stationary characteristics of viewing time sequences without requiring explicit assumptions about the data distribution.

\subsection{Evaluation Protocol and Metric}
All models are evaluated under a multi-step recursive forecasting setting combined with a rolling-origin evaluation protocol. For formal evaluation, we extend the general notation $\hat{y}_t$ to $\hat{y}_{u,w,t}$, which represents the predicted viewing time for user $u$ within the $w$-th rolling window at time index $t \in \{w + N, \dots, w + N + H - 1\}$. This notation maps the general model outputs to specific experimental instances. For each user, a training window of fixed length $N$ is slid along the sequence with a step size of one. At each window $w$, the model is trained on current observations to generate recursive predictions $\{\hat{y}_{u,w,t}\}$. Previously predicted values are then used as inputs for subsequent steps without accessing any ground-truth future observations. Prediction accuracy is evaluated using the Root Mean Square Error (RMSE), computed independently for each predicted time index $t$ as:

\begin{equation}
\text{RMSE}_t = \sqrt{\frac{1}{UW} \sum_{u=1}^{U} \sum_{w=1}^{W} (\hat{y}_{u,w,t} - y_{u,w,t})^2},
\end{equation} where $U$ denotes the number of users and $W$ is the number of rolling windows per user.

Fig.~\ref{fig:rmse_all} illustrates the RMSE comparison of five forecasting methods as the training size varies from $N = 10$ to $N = 80$. Overall, the results indicate that forecasting performance is strongly influenced by the training size when $N$ is small, whereas the differences among methods become more stable once sufficient historical data are available. 

When $N = 10$, the AR model exhibits clear instability, with a sharp increase in RMSE at the final prediction steps. This behavior suggests severe error accumulation under limited training data. In contrast, the remaining methods maintain RMSE values within a relatively stable range. Among them, Auto-ARIMA achieves the lowest and most stable RMSE across all prediction steps. Linear Regression shows a nearly monotonic increase in RMSE as the prediction horizon extends and reaches the highest values among the models whose errors remain controlled. Both SVR and DTR produce higher RMSE than Auto-ARIMA. For $N = 20$, AR continues to generate higher errors than the other methods, particularly at distant prediction steps where the RMSE increases rapidly. Linear Regression maintains its increasing error trend with respect to the prediction horizon, reflecting performance degradation in long-term forecasting. Meanwhile, Auto-ARIMA remains stable and achieves the best overall performance under this configuration. 

From $N = 30$ onward, the forecasting behavior changes noticeably. The monotonic increase of RMSE with respect to the prediction step largely disappears, and the errors fluctuate within a relatively narrow range. Under these configurations, both AR and Auto-ARIMA frequently achieve the lowest RMSE values (approximately $6.0$--$8.5$). SVR consistently yields higher RMSE than AR and Auto-ARIMA, while DTR generally produces the largest errors and exhibits considerable variability across prediction steps. As the training size further increases from $N = 50$ to $N = 80$, the RMSE values of all methods remain relatively stable and do not decrease significantly. This observation indicates that increasing the training size beyond a certain threshold (around $N \geq 30$) does not provide substantial improvement in forecasting accuracy for this highly variable dataset. The relative ranking of the methods also remains largely unchanged in this region. Overall, Auto-ARIMA demonstrates the most stable and consistent performance across all experimental configurations. The AR model achieves competitive accuracy when sufficient training data are available, but it becomes unreliable when the sample size is too small. SVR and DTR generally produce higher RMSE, with DTR exhibiting the weakest performance among the five compared methods. These findings highlight the importance of model stability and robustness when handling short and highly dynamic time series.

% \subsection{Discussion}

% \section{Chunk-based Preloading Algoirthm}\label{sec:chunk_loading_alg}

\section{Conclusion}\label{sec:conclusion}
In this paper, we built and publicly released a dataset of user viewing times for 100 short videos across different categories. Based on this dataset, we compared five forecasting methods, including LR, AR, Auto-ARIMA, DTR, and SVR, to predict user viewing time in short-video streaming. The results show that Auto-ARIMA achieves the lowest and most stable forecasting errors across most experimental settings. In contrast, the AR model becomes unstable when the training size is small, with RMSE increasing sharply at longer prediction steps. Linear Regression remains relatively stable but generally produces higher errors than Auto-ARIMA. SVR and especially DTR usually yield larger prediction errors in most cases. These findings highlight the importance of selecting an appropriate forecasting model. More accurate viewing-time prediction can help reduce unnecessary data preloading and improve overall streaming efficiency. In future work, we plan to explore more advanced forecasting models and adaptive preloading strategies to further improve system performance.

\bibliographystyle{IEEEbib}
\bibliography{reference}

\end{document}